\documentclass[aps,prl,showpacs,twocolumn]{revtex4}
\ifx\pdftexversion\undefined
\usepackage[dvips]{graphics}
\else
\usepackage[pdftex]{graphics}
\fi

\usepackage{graphics,epsfig}

\begin{document}

\title{How to play two-players restricted quantum games with 10 cards}

\author{Diederik Aerts$^1$, Haroun Amira$^1$, Bart D'Hooghe$^1$, Andrzej 
Posiewnik$^2$ and Jaroslaw Pykacz$^3$}
\affiliation{$^1$Center Leo Apostel for Interdisciplinary Studies and Department of Mathematics, Vrije Universiteit Brussels, 1160 
Brussels, Belgium\\
$^2$Institute of Theoretical Physics and Astrophysics, University of 
Gda\'{n}sk, 80-952 Gda\'{n}sk, Poland\\
$^3$Institute of Mathematics, University of Gda\'{n}sk, 80-952 
Gda\'{n}sk, Poland}

\email{diraerts@vub.ac.be, Haroun.Amira@vub.ac.be, bdhooghe@vub.ac.be, fizap@julia.univ.gda.pl, pykacz@math.univ.gda.pl}

\date{\today}

\begin{abstract}
We show that it is perfectly possible to play `restricted' two-players,
two-strategies quantum games proposed originally by Marinatto and Weber 
\cite{MW00} having as the only equipment a pack of 10 cards. The 
`quantum board'
of such a model of these quantum games is an extreme simplification of
`macroscopic quantum machines' proposed by Aerts et al. in numerous 
papers 
\cite{Aerts86,Aer91,Aerts93,Aeretal00} that allow to simulate by
macroscopic means various experiments performed on two entangled 
quantum objects.
\end{abstract}

\maketitle

\section{Introduction}

Although the theory of quantum games, originated in 1999 by Meyer 
\cite{Mey99}
and Eisert, Wilkens, and Lewenstein \cite{EWL99} is only six years old,
numerous results obtained during these years \cite{PS03} have shown that
extending the classical theory of games to the quantum domain opens new
interesting possibilities. Although Eisert and Wilkens \cite{EW00} noticed 
that
\textquotedblleft Any quantum system which can be manipulated by two 
parties
or more and where the utility of the moves can be reasonably 
quantified, may
be conceived as a quantum game\textquotedblright , the extreme fragility of
quantum systems may make playing quantum games difficult. In this 
respect it
is interesting whether quantum games with all their `genuine quantum'
features could be played with the use of suitably designed macroscopic
devices. The aim of this letter is to show that this is possible, at 
least
in the case of a `restricted' version of a two-players, two-strategies
quantum game proposed by Marinatto and Weber \cite{MW00} in which only
identity and spin-flip operators are used. Moreover, we show that this 
can
be done at once by anyone equipped with a pack of 10 cards bearing 
numbers $%
0,1,...,9$.

Our idea of playing quantum games with macroscopic devices stems from
the invention devices proposed by one of us \cite{Aer91} 
that
perfectly simulate the behavior and measurements performed on two maximally
entangled spin-1/2 particles. For example, they allow to violate the 
Bell
inequality with $2\sqrt{2}$, exactly `in the same way' as it is 
violated in
the EPR experiments. A more recent and further elaborated model 
consists of
two coupled spin-1/2 for which measurements are defined using `randomly
breaking measurement elastics' \cite{Aerts93,Aeretal00}. In this paper 
we
use the older model for a single spin-1/2 for which measurements are 
defined
using `randomly selected measurement charges' \cite{Aerts86,Aer91}. 
In order to play Marinatto and Weber's `restricted' version of
two-players, two-strategies quantum game we shall not use the `full 
power'
of this machine, but we give its complete description such that the principle of what we try to do is clear.

\section{Macroscopic simulations of Marinatto and Weber's quantum 
games}

\subsection{The quantum machine}

The quantum machine is a model for a spin-1/2 particle consisting of a point
particle with negative charge $q$ on the surface $S^{2}$ of a 
3-dimensional
unit sphere \cite{Aerts86,Aerts93}. The spin-state ${|\psi \rangle 
}=\left(
\cos {\frac{\theta }{2}}e^{\frac{-i\phi }{2}},\sin {\frac{\theta 
}{2}}e^{%
\frac{i\phi }{2}}\right) $ is represented by the point $v(1,\theta 
,\phi )$
on $S^{2}$. All points of the sphere represent states of the spin: 
points on
the surface $S^{2}$ correspond to pure states, interior points 
$v(r,\theta
,\phi )$ represent mixed states ${|\psi \rangle }{\langle \psi 
|=\frac{1}{2}}%
\left( 
\begin{array}{cc}
1+r\cos \theta  & r\sin \theta e^{-i\phi } \\ 
r\sin \theta e^{i\phi } & 1-r\cos \theta 
\end{array}%
\right) $, such that the point $v(0,\theta ,\phi )$ in the center of 
the
sphere represents the density matrix $\left( 
\begin{array}{cc}
\frac{1}{2} & 0 \\ 
0 & \frac{1}{2}%
\end{array}%
\right) $. Hence states are represented equivalently as this is the case in the Bloch model for the spin 1/2.

\begin{figure}[tbh]
\begin{center}
\includegraphics{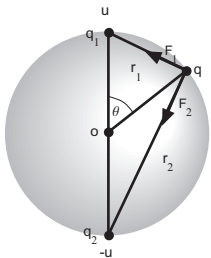}   
\end{center}
\caption{The macroscopic quantum machine}
\label{fig:mqg01}
\end{figure}

A measurement $\alpha _{u(\theta ,\phi )}$ along the 
direction $u$
consists in placing a positive charge $q_{1}$ in $u$ and a positive 
charge $%
q_{2}$ in $-u$. The charges $q_{1}$ and $q_{2}$ are taken at random 
from the
interval $[0,Q]$ and their distribution within this interval is assumed 
to
be uniform, but they have to satisfy the constraint $q_{1}+q_{2}=Q.$ So 
in
fact we can think that only $q_{1}$ is taken at random from the 
interval $%
[0,Q]$ and that $q_{2}=Q-q_{1}$. If the initial state of the machine is 
as
depicted on Fig.~\ref{fig:mqg01}, the forces $F_{1}$ and $F_{2}$ between 
the
negative charge $q$ and, respectively, positive charges $q_{1}$ and 
$q_{2}$
are 
\begin{equation}
F_{1}=C\frac{qq_{1}}{|r_{1}|^{2}}\text{ \qquad and\qquad 
}F_{2}=C\frac{qq_{2}%
}{|r_{2}|^{2}}
\end{equation}%
If $F_{1}>F_{2}$ the electromagnetic forces pull the particle to the 
point $u
$ where it stays and the measurement is said to yield outcome `spin 
up', and
if $F_{1}<F_{2}$ the particle is pulled to $-u$ yielding outcome `spin
down'. Denoting the angle between directions $v$ and $u$ by $\theta $, 
one
obtains $r_{1}=2\sin \frac{\theta }{2}$ and $r_{2}=2\cos \frac{\theta }{2}.$ Hence the probability that 
$F_{1}>F_{2}$
is: 
\begin{eqnarray}
P\left( C\frac{qq_{1}}{|r_{1}|^{2}}>C\frac{qq_{2}}{|r_{2}|^{2}}\right) 
&=&P\left( q_{1}>Q\sin ^{2}\frac{\theta }{2}\right) 
\end{eqnarray}%
which, since $q_{1}$ is assumed to be uniformly distributed in the 
interval $%
[0,Q]$, yields 
\begin{eqnarray}
P(\text{spin up}) 
&=&\frac{Q-Q\sin ^{2}\frac{\theta }{2}}{Q}=\cos 
^{2} 
\frac{\theta }{2}
\end{eqnarray}%
and similarly 
\begin{equation}
P(\text{spin down})=P(F_{1}<F_{2})=\sin ^{2}\frac{\theta 
}{2} 
\end{equation}%
which coincides with the quantum mechanical probability distribution 
over
the set of outcomes for a spin-1/2 experiment.

A macroscopic model for a quantum system of two entangled spin-1/2 
particles
in the singlet state \cite{Aer91} can be constructed by `coupling' two 
such
sphere models by adding a rigid but extendable rod with a fixed center 
that
connects negative charges representing `single' particles (Fig.~\ref%
{fig:mqg02}). Because of this rod the two negative charges are 
`entangled'
since a measurement performed on one of them necessarily influences the
state of the other one.

\subsection{Quantum games proposed by Marinatto and Weber}

The `restricted' version of two-players, two-strategies quantum games
proposed by Marinatto and Weber is as follows: The `quantum board' of 
the
game consists of two qubits that are in a definite initial state 
(entangled
or not). Each of two players obtains one qubit and his/her strategy 
consists
in applying to it either the identity or the spin-flip operator, or a
probabilistic mixture of both. Then the state of both qubits is 
measured and
the players get their payoff calculated according to the specific 
bimatrix
of the played game and the results of measurements. Marinatto and Weber 
in
their paper \cite{MW00} considered a game with a payoff bimatrix:

\begin{equation}
\begin{array}{ccc}
& \text{Bob: \emph{O}} & \text{Bob: \emph{T}} \\ 
\text{Alice: \emph{O}} & (\alpha ,\beta ) & (\gamma ,\gamma ) \\ 
\text{Alice: \emph{T}} & (\gamma ,\gamma ) & (\beta ,\alpha )%
\end{array}
\label{BoS}
\end{equation}
which, if $\alpha >\beta >\gamma $, is the payoff bimatrix of the 
Battle of
the Sexes game (Alice wants to go to the Opera while Bob prefers to 
watch
Television, so if they both choose \emph{O} Alice's payoff 
$\$_A(O,O)=\alpha 
$ is bigger than Bob's payoff $\$_B(O,O)=\beta $, and if they both 
choose 
\emph{T} their payoffs are the opposite. Since they both prefer to stay
together, if their strategies mismatch they are both unhappy and get 
the
lowest payoff $\gamma $). Marinatto and Weber showed that if the 
initial
state of the pair of qubits is not entangled, the quantum version of 
the
game reproduces exactly the classical Battle of the Sexes game played 
with
mixed strategies, but if the game begins with an entangled state of the
`quantum board': $\mid \psi _{in}\rangle =a|OO\rangle +b|TT\rangle $, 
$%
|a|^2+|b|^2=1$, then the expected payoff functions for both players
crucially depend on the values of squared moduli of `entanglement
coefficients' $|a|^2$ and $|b|^2$, and allow for new `solutions' of the 
game
not attainable in the classical or factorizable quantum case.

\subsection{Marinatto and Weber's `restricted' quantum game realized by the macroscopic quantum machine}

\begin{figure}[htb!]
\begin{center}
\includegraphics{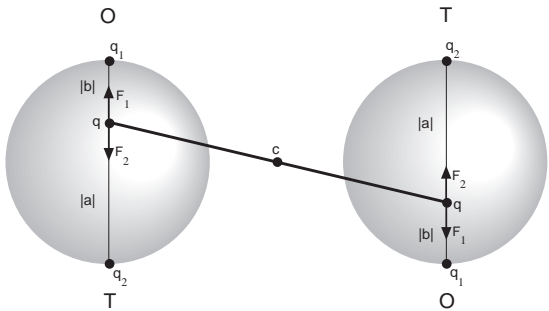}   
\end{center}
\caption{The initial general `entangled' state of Aerts' quantum machine.}
\label{fig:mqg02}
\end{figure}

Let us look now how simply Marinatto and Weber's `restricted' quantum 
game
can be macroscopically realized with the use of the macroscopic quantum machine. We
describe firstly the macroscopic realization of the game that begins 
with a
general entangled state 
\begin{equation}
\mid \psi _{in}\rangle =a|OO\rangle +b|TT\rangle ,\text{ \qquad }%
|a|^{2}+|b|^{2}=1.  \label{in}
\end{equation}%
The game that begins with a non-entangled state can be obtained from it 
as a
limit in which either $|a|^{2}=0$ or $|b|^{2}=0$. The initial 
configuration
of the macroscopic machine that realizes the state (\ref{in}) is 
depicted on
Fig.~\ref{fig:mqg02}. 

Applying the spin-flip operator by any of the players is realized as
exchanging the labels \emph{O} and \emph{T} on his/her sphere. Let us 
note
that this is a local operation since it does not influence in any way 
the
sphere of the other player. Applying the identity operator obviously 
means
doing nothing. When both players make (or not) their movements, the
measurement is performed which, similarly to the original Aerts' 
proposal in 
\cite{Aer91}, consists in placing a positive charge $q_{1}$ on the 
North
pole and a positive charge $q_{2}$ on the South pole of the Alice's 
sphere,
and the same charges, respectively, on the South and North poles of the
Bob's sphere (i.e., on the Bob's sphere $q_{1}$ is placed on the South 
pole
and $q_{2}$ on the North pole). Again, charges $q_{1}$ and $q_{2}$ are 
taken
at random from the interval $[0,Q]$ with uniform distribution 
satisfying the
constraint $q_{1}+q_{2}=Q.$ Assuming for simplicity that forces between
`left' positive and `right' negative, resp. `right' positive and `left'
negative charges are negligible (which can be achieved by using a rod 
that
is long enough or by suitable screening) we can make analogous 
calculations
as for the single sphere model. The forces $F_{1}$ and $F_{2}$ between 
the
negative charges $q$ placed at both ends of the rod and, respectively,
positive charges $q_{1}$ and $q_{2}$ are 
\begin{equation}
F_{1}=C\frac{qq_{1}}{|b|^{2}}\text{ \qquad and\qquad 
}F_{2}=C\frac{qq_{2}}{%
|a|^{2}}
\end{equation}%
The final state of the machine (the result of measurement) depends on 
which
force, $F_{1}$ or $F_{2}$, is bigger. If the labels $O$ and $T$ are 
placed
as on Fig.~\ref{fig:mqg02}, the result of the measurement is $(O,O)$ 
iff $%
F_{1}>F_{2}$, and $(T,T)$ iff $F_{1}<F_{2}$. The probability that $%
F_{1}>F_{2}$ is as follows: 
\begin{equation}
P\left( F_{1}>F_{2}\right) =P\left( q_{1}|a|^{2}>q_{2}|b|^{2}\right)
=P\left( q_{1}>Q|b|^{2}\right)   \label{prob1}
\end{equation}%
which, since $q_{1}$ is assumed to be uniformly distributed in the 
interval $%
[0,Q]$, yields 
\begin{equation}
P(O,O)=P(F_{1}>F_{2})=\frac{Q-Q|b|^{2}}{Q}=1-|b|^{2}=|a|^{2}.  
\label{prob2}
\end{equation}%
Of course in this case 
\begin{equation}
P(T,T)=P(F_{1}<F_{2})=1-|a|^{2}=|b|^{2}.  \label{prob3}
\end{equation}

Let us assume, following Marinatto and Weber, that Alice applies the
identity operator (in our model: undertakes no action) with probability 
$p$
and applies the spin-flip operator (in our model: exchanges the labels 
$O$
and $T$ on her sphere) with probability $1-p$, and Bob does the same on 
his
side with respective probabilities $q$ and $1-q$. Consequently, when 
both
players make (or not) their movements, the configuration depicted on 
Fig.~%
\ref{fig:mqg02} occurs with probability $pq$, and the result of the
measurement is $(O,O) $ with probability $pq|a|^{2}$ and $(T,T)$ with
probability $pq|b|^{2}$. Taking into account three other possibilities
(Alice undertaking no action and Bob exchanging the labels, Alice 
exchanging
the labels and Bob undertaking no action, and both of them exchanging 
their
labels) which occur with respective probabilities $p(1-q)$, $(1-p)q$, 
and $%
(1-p)(1-q)$, and the payoff bimatrix (\ref{BoS}), we obtain the 
following
formulas for the expected payoff of Alice:

\begin{equation}
\begin{array}{ll}
\overline{\$}_{A}(p,q) & =pq(|a|^{2}\alpha +|b|^{2}\beta )+p(1-q)\gamma 
\\ 
& \text{ }+(1-p)q\gamma +(1-p)(1-q)(|a|^{2}\beta +|b|^{2}\alpha ) \\ 
& =p[q(\alpha +\beta -2\gamma )-\alpha |b|^{2}-\beta |a|^{2}+\gamma ] 
\\ 
& \text{ }+q(-\alpha |b|^{2}-\beta |a|^{2}+\gamma )+\alpha 
|b|^{2}+\beta
|a|^{2},%
\end{array}
\label{payoffAlice}
\end{equation}%
and the expected payoff of Bob: 
\begin{equation}
\begin{array}{ll}
\overline{\$}_{B}(p,q) & =pq(|b|^{2}\alpha +|a|^{2}\beta )+p(1-q)\gamma 
\\ 
& \text{ }+(1-p)q\gamma +(1-p)(1-q)(|a|^{2}\alpha +|b|^{2}\beta ) \\ 
& =q[p(\alpha +\beta -2\gamma )-\alpha |a|^{2}-\beta |b|^{2}+\gamma ] 
\\ 
& \text{ }+p(-\alpha |a|^{2}-\beta |b|^{2}+\gamma )+\alpha 
|a|^{2}+\beta
|b|^{2}.%
\end{array}
\label{payoffBob}
\end{equation}%
Let us note that these formulas, although obtained from the 
`mechanistic'
model through `classical' calculations are \emph{exactly }the same as
formulas (7.3) of Marinatto and Weber \cite{MW00} for the payoff 
functions
of Alice and Bob in their `reduced' version of the quantum Battle of 
the
Sexes game that begins with a general entangled state (\ref{in}).

The macroscopic model of the quantum game that begins with a 
non-entangled
state $\mid \psi _{in}\rangle =|OO\rangle $ can be obtained by putting 
in (%
\ref{in}) $a=1$ and $b=0$, which means that in this case the rod on 
Fig.~\ref%
{fig:mqg02} leads from the North pole of Alice's sphere to the South 
pole
of Bob's sphere. In this case we obtain 
\begin{equation}
\begin{array}{ll}
\overline{\$}_{A}(p,q) & =p[q(\alpha +\beta -2\gamma )+\gamma -\beta ] 
\\ 
& \text{ }+q(\gamma -\beta )+\beta , \\ 
\overline{\$}_{B}(p,q) & =q[p(\alpha +\beta -2\gamma )-\alpha 
|a|^{2}-\beta
|b|^{2}+\gamma ] \\ 
& \text{ }+p(\gamma -\beta )+\alpha ,%
\end{array}
\label{payoffNonent}
\end{equation}%
again in the perfect agreement with Marinatto and Weber's \cite{MW00}
formulas (3.3).

This result might be surprising since the rod connecting two particles
represents entanglement in the macroscopic quantum machine so one could expect that 
when
the initial state of the game is not entangled, this connection should 
be
broken. However, it should be noticed that in the device depicted on 
Fig.~%
\ref{fig:mqg02} the rod connecting two particles is, in fact, 
redundant.
The reason for which we left it on Fig.~\ref{fig:mqg02} is twofold:
firstly, we wanted to stress that our idea of a macroscopic device that
allows to play quantum games stems from the ideas published in \cite%
{Aer91,Aeretal00}, and secondly, this rod will be essential for 
macroscopic
simulations of other quantum games, more general than Marinatto and 
Weber's
`restricted' ones.

Thus, we see that what vanishes in the `non-entanglement' limit of the
considered quantum game is the `randomness in measurement', since now
(except for the zero-probability case when $q_{1}=0$, $q_{2}=Q$) the 
initial
state of the machine does not change in the course of the measurement
whatever is the value of $q_{1}$.

\subsection{Marinatto and Weber's `restricted' quantum game realized 
with a
pack of 10 cards}

\begin{figure}[htb!]
\begin{center}
\includegraphics{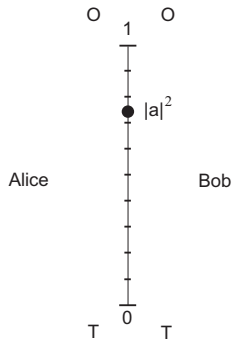}   
\end{center}
\caption{The board to play `restricted' quantum games with 10 cards.}
\label{fig:mqg03}
\end{figure}

The lack of any importance of the connecting rod and the fact that all
distances, charges, and forces in the device depicted on Fig.~\ref%
{fig:mqg02} are symmetric with respect to the middle of the rod allow 
to
produce a still more simple model of the considered game, in fact so 
simple
that it can be played with a piece of paper and a pack of 10 cards 
bearing
numbers $0,1,...,9$. The game is played in three steps. In the first 
step
the initial `quantum' state of the game (\ref{in}) is fixed. Since only 
the
squared moduli of entanglement coefficients $|a|^{2}$ and $|b|^{2}$ are
important and $|a|^{2}+|b|^{2}=1$, it is enough to fix a point 
representing $%
|a|^{2}$ in the interval $[0,1]$ (Fig.~\ref{fig:mqg03}). 

In the next step the players exchange, or not, labels $O$ and $T$ on 
their
sides modelling in this way application of spin-flip, resp. identity,
operators. In the third step a measurement is made, which is executed 
by
choosing at random a number in the interval $[0,1]$. If a chosen number 
is
smaller than $|a|^{2}$ which, if the probability distribution is 
uniform in $%
[0,1]$, happens with the probability $|a|^{2}$, the result of the
measurement is given by labels placed by both players close to $1$,
otherwise by labels placed close to $0$. Although random choosing of a
number may be executed in many ways, we propose to use a pack of 10 
cards
bearing numbers $0,1,...,9$ which allows to draw one by one, with 
uniform
probability, consecutive decimal digits of a number until we are sure 
that
the emerging number is either definitely bigger or definitely smaller 
than $%
|a|^{2}$(we put aside the problem of drawing in this way the number 
that 
\emph{exactly} equals $|a|^{2}$ since its probability is $0$, as well 
as the
fact that in a series of $n$ drawings we actually choose one of 
$10^{n}$
numbers represented by separate points uniformly distributed in the 
interval 
$[0,1-10^{-n}]$). Of course calculations of the payoff functions that 
we
made while describing the device depicted on Fig.~\ref{fig:mqg02} are
still valid in this case, so we again obtain perfect macroscopic 
simulation
of Marinatto and Weber's `restricted' two-players, two-strategies 
quantum
games.

Thus, one does not have to be equipped with sophisticated and costly 
devices
and perform subtle manipulations on highly fragile single quantum 
objects in
order to play quantum games, at least in the `restricted' Marinatto and
Weber's version: all that suffices is a piece of paper and a pack of 10
cards!

\noindent \textbf{Acknowledgments} This work was carried out within the
projects G.0335.02 and G.0452.04 of the Flemish Fund for Scientific research.


\begin{thebibliography}{99}
\bibitem{MW00} L. Marinatto and T. Weber, Phys. Lett. A, \textbf{272}, 
291
(2000).

\bibitem{Aerts86} D. Aerts, J. Math. Phys., \textbf{27}, 202 (1986).

\bibitem{Aer91} D. Aerts, Helv. Phys. Acta, \textbf{64}, 1 (1991).

\bibitem{Aerts93} D. Aerts, Int. J. Theor. Phys., \textbf{32}, 2207 
(1993).

\bibitem{Aeretal00} D. Aerts, S. Aerts, J. Broekaert and L. Gabora,
Foundations of Physics, \textbf{30}, 1387 (2000).

\bibitem{Mey99} D. A. Meyer, Phys. Rev. Lett.\textbf{\ 82}, 1052 
(1999).

\bibitem{EWL99} J. Eisert, M. Wilkens, and M. Lewenstein, Phys. Rev. 
Lett. 
\textbf{83}, 3077 (1999).

\bibitem{PS03} Brief introduction into quantum games theory can be 
found
in E. W. Piotrowski and J. S\l adkowski e-print quant-ph/0308027.

\bibitem{EW00} J. Eisert and M. Wilkens, J. Modern Optics, \textbf{47}, 
2543
(2000).

\end{thebibliography}
\end{document}